\documentclass[twocolumn,showpacs]{revtex4}

\newcommand{\ltsim}{\mbox{{\raisebox{-0.4ex}{$\stackrel{<}{{\scriptstyle\sim
}}
$}}}}

\def\khg{$\alpha$-(BEDT-TTF)$_2$KHg(SCN)$_4$}
\def\nh4{$\alpha$-(BEDT-TTF)$_2$NH$_4$Hg(SCN)$_4$}

\usepackage{graphicx}
\usepackage{dcolumn}
\usepackage{amsmath}
\begin{document}

\title{A new quantum fluid at high magnetic 
fields in the marginal charge-density-wave
system $\alpha$-(BEDT-TTF)$_2M$Hg(SCN)$_4$
(where $M=$~K and Rb)}

\author{N.~Harrison$^1$, J.~Singleton$^1$, A. Bangura$^2$,
A.~Ardavan$^2$, P.A. Goddard$^{1,2}$, R.D. McDonald$^1$
and L.K. Montgomery$^3$
}

\affiliation{
$^1$National High Magnetic Field Laboratory, LANL, 
MS-E536, Los~Alamos, New~Mexico~87545, USA\\
$^2$The Clarendon Laboratory, Parks Road, 
Oxford~OX1~3PU, United Kingdom\\
$^3$Department of Chemistry, 
Indiana University, Bloomington, IN~47405, USA}

\begin{abstract}
Single crystals of the organic charge-transfer salts
$\alpha$-(BEDT-TTF)$_2M$Hg(SCN)$_4$ 
have been studied using Hall-potential measurements
($M=$K) and magnetization experiments ($M=$~K, Rb).
The data show that two types of 
screening currents occur within
the high-field, low-temperature CDW$_x$ phases
of these salts
in response to time-dependent magnetic fields.
The first, which gives rise to the induced Hall 
potential, is a free current (${\bf j}_{\rm free}$),
present at the surface of the sample. 
The time constant for the decay of these currents
is much longer than that expected from the sample
resistivity. The second component of the current 
appears to be magnetic (${\bf j}_{\rm mag}$), 
in that it is a microscopic, quasi-orbital
effect; it is evenly distributed within
the bulk of the sample upon saturation.
To explain these data, we propose a 
simple model invoking a new
type of quantum fluid comprising a CDW coexisting
with a two-dimensional Fermi-surface pocket
which describes the two types of current.
The model and data are able to account
for the body of previous experimental 
data which had generated apparently contradictory
interpretations in 
terms of the quantum Hall effect
or superconductivity. 
\end{abstract}

\pacs{71.45.Lr, 71.20.Ps, 71.18.+y}

\maketitle

\section{introduction}
In their rudimentary form, charge-density waves (CDWs) consitute a
simple one-dimensional spin singlet groundstate in which the
collective mode consists of a charge density modulated at a
characteristic wavevector~$Q=2k_{\rm F}$.
Here, $k_{\rm F}$ is the Fermi wavevector of
the Fermi-surface section responsible for CDW 
formation~\cite{gruner1,gruner2,solyom1}.  
Compared to superconductors, their
behavior in a magnetic field is relatively 
easy to predict, providing
perhaps the simplest example of a 
singlet ground state reaching the
Pauli limit~\cite{dietrich1,zanchi1,mckenzie1,harrison1}.  
This limit is defined as the magnetic field 
required to cause the energy of the
partially spin-polarized normal state to 
become lower than that of the condensate.  

In spite of the simplicity of CDW systems, 
the large values of the CDW condensation 
temperatures $T_{\rm p}$, typically of 
order 100~K~\cite{gruner1,gruner2}, 
make the Pauli limit inaccessible in
standard laboratory magnetic fields for 
the vast majority of CDW 
materials~\cite{balls}.
Our reason for studying the title compounds,
$\alpha$-(BEDT-TTF)$_2M$Hg(SCN)$_4$, 
is that they lie at the far lower end of the
spectrum of transition temperatures, with 
$T_{\rm p}\approx 8$~K~\cite{foury1} for $M=$~K 
and $T_{\rm p}\approx~12$~K for $M=$~Rb, 
making them perhaps the most marginal of CDW systems.
Consequently, they are the only CDW compounds 
in which the gap is
sufficiently low for the primary CDW phase, 
CDW$_0$, to have been
shown to be Pauli limited by a relatively 
modest magnetic field of $\mu_0H\approx 23$~T for 
$M=$~K and $\mu_0H \approx 32$~T 
for $M=$~Rb~\cite{mckenzie1,harrison1,biskup1,christ1,harrison2}.

\begin{figure}[htbp]
\centering
\includegraphics[width=8cm]{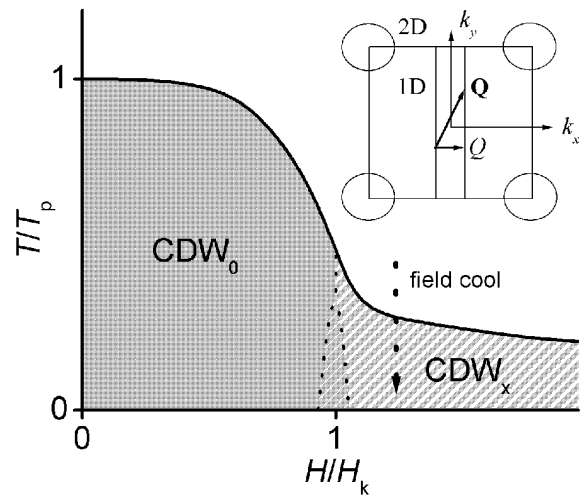}
\caption{Charge-density Wave formation in
$\alpha$-(BEDT-TTF)$_2M$Hg(SCN)$_4$ salts.  
A schematic of the phase diagram as a 
function of reduced temperature $T/T_{\rm p}$ and reduced
magnetic field $H/H_{\rm k}$ (based on Ref.~\cite{harrison8}).
The second order transition temperature
$T_{\rm p}$ at $H=$~0 is approximately 8~K 
in the $M=$~K salt and 12~K in the $M=$~Rb salt.  
Likewise, the first order ``kink''
transition field $\mu_0H$ at $T=$~0 
(roughly corresponding to the
Pauli paramagnetic limit), is 23~T for $M=$K
and 32~T for $M=$Rb. `CDW$_0$' refers
to the proposed conventional CDW phase 
while `CDW$_x$' refers to an incommensurate phase 
that comes into existence at high magnetic
fields.  The inset shows a simplified schematic 
of the Fermi-surface in the repeated Brillouin 
zone representation (after Ref.~\cite{singleton1}) 
which consists of open one-dimensional
(1D) sections which run along $k_y$
and a closed two-dimensional (2D) section.
(An accurate representation of the 
actual Fermi surface topolgy is given in Ref.~\cite{harrison3}.)
CDW formation is thought to occur for a 
characteristic ``nesting vector'' ${\bf Q}$~\cite{harrison3}
(with a component $Q$ along $k_x$), 
causing the 1D section to become gapped.  }
\label{Fermisurface}
\end{figure}

It must be stated, however, that 
the marginal CDW properties of
$\alpha$-(BEDT-TTF)$_2M$Hg(SCN)$_4$ render
it somewhat beyond the predictive
power of the standard theory~\cite{dietrich1}.  
For instance, the low
transition temperature results in 
a CDW with a weak charge modulation
that is vulnerable to fluctuations 
and sample-dependent effects~\cite{foury1}.  
Furthermore, the presence of large sections of
the Fermi surface left ungapped by 
the CDW order~\cite{harrison3} (see 
inset to Fig.~\ref{Fermisurface}) 
causes the energy associated with
the coupling of the magnetic field to 
orbital degrees of freedom of
the itinerant electrons to rival the condensation energy.
This leads to a modification of the quantum oscillatory
effects~\cite{harrison1,harrison4,harrison5} 
and to the possibility of
field-induced CDW sub-phases~\cite{zanchi1,andres1}.

At magnetic fields above the Pauli paramagnetic limit,
$\alpha$-(BEDT-TTF)$_2M$Hg(SCN)$_4$ then crosses 
over into a new regime in which the transition 
temparature becomes even lower~\cite{biskup1,christ1,harrison2}, 
falling to between 2 and 4~K:
a schematic of the phase diagram is shown in 
Fig.~\ref{Fermisurface}.
Since the conventional CDW phase, 
CDW$_0$, is no longer stable at high
magnetic fields, the spin-up and spin-down 
electrons must instead form
independent charge and spin 
modulations with momentum vectors $Q_\uparrow$ and 
$Q_\downarrow$~\cite{zanchi1,mckenzie1}, leading to a
different type of CDW phase.  

Experimental studies of this high-magnetic-field phase, 
denoted CDW$_x$, 
lead one to question whether it is a CDW at all.  
While conventional wisdom has it that 
CDWs constitute a class of narrow-gap 
insulators~\cite{gruner1,gruner2},
experimental studies find behavior that is
reminiscent either of the quantum Hall
effect~\cite{harrison6,hill1,harrison7,honold1,honold2}
or superconductivity~\cite{harrison2,harrison8,harrison9,mielke1}.
These experimental findings include
a sharp drop in the electrical resistivity at 
low temperatures, persistent currents and unusual 
Hall-voltage phenomena.  Problems arise because the quantum Hall
effect~\cite{chakraborty1} and superconductivity~\cite{tinkham1}
have fundamentally different origins.  
If one attempts to categorize interpretations 
of experimental data in terms of the
quantum Hall effect~\cite{singleton1} 
or superconductivity~\cite{harrison8}, 
one obtains two sets of conclusions that are in contradiction.

The present paper seeks to reconcile the body of experimental 
evidence~\cite{harrison2,harrison6,hill1,harrison7,honold1,
honold2,harrison8,harrison9,mielke1} 
in exploring the predicted properties of
a new type of quantum fluid,
consisting of a CDW coexisting with a 2D Fermi surface.
A simple model shows that the
exchange of quasiparticles between these subsystems
results in the sample partially
screening time- and spatially-varying
electromagnetic fields via two types of current flow;
this result is supported by the
experimental data in this work, which show that
the screening currents observed in earlier magnetic
measurements~\cite{harrison2,harrison6,harrison9,mielke1,harrison12}
comprise two parts.  
The first is weighted mostly towards the sample edges
and can be considered a free current 
${\bf j}_{\rm free}$, furnishing a Hall-potential gradient. 
The second contribution is analogous to the
``magnetic currents'' ${\bf j}_{\rm mag}$ used to represent
the effects of localised orbital moments~\cite{bleaney} in
producing magnetization.
These currents, which are essentially of infinite
duration~\cite{harrison2}, represent a local, microscopic current
flow within the bulk of the sample;
they occur if the surface scrrening mechanism
fails, causing the CDW, instead, to
be coerced by an external change in magnetic field into a
non-equilibrium state.  In this case, bulk currents
are supported by CDW pinning forces and are not free.

The remainder of this paper is organized as follows.
Experimental details are given in Section~\ref{experiment}, whilst
Section~\ref{honoldstuff} describes the Hall potential
measurements which characterize the surface current density
${\bf j}_{\rm free}$. 
The ``magnetic'' current density ${\bf j}_{\rm mag}$
is studied using torque magnetometry in Section~\ref{torquetorque}.
Conclusions are given in Section~\ref{discussion};
for ease of reference, the model describing the 
microscopic mechanism that leads to ${\bf j}_{\rm free}$ and
${\bf j}_{\rm mag}$ is described in the Appendix.

\section{experimental}
\label{experiment}
The free electrical current distribution, ${\bf j}_{\rm free}$, was
determined using the variant of the 
Corbino geometry described in Ref.~\cite{honold2} in which pairs of 
small graphite-paint contacts are placed on 
the outer edge and upper surface of
a \khg ~sample. 
Currents were induced in the sample by applying
a small sinusoidal oscillatory field of amplitude $\mu_0 \tilde{H}$ and
angular frequency $\omega$ superimposed on the quasistatic field
$\mu_0H$ provided by a 33 T Bitter coil at NHMFL, Tallahassee.
Alternatively, currents were induced by
sweeping the quasistatic field at a rate
$\mu_0(\partial H/\partial t)$.
Provided that the sample dimensions are
much smaller than its skin depth (so that it is entirely
penetrated by the changing field),
the resulting Hall potential between the edge of the sample, of area $A$,
and its geometrical centre is~\cite{honold2}
\begin{equation}
V_{\rm H} = \frac{A}{4\pi}
\left(\frac{\rho_{xy}}{\rho_{||}}\right) \mu_0 \frac{\partial H}{\partial t}
~~~{\rm (swept~field);}
\label{hallpotential2}
\end{equation}
and
\begin{equation}
V_{\rm H} = \frac{A}{4\pi}\left(\frac{\rho_{xy}}{\rho_{||}}\right)
\omega \mu_0 \tilde{H}~~~{\rm (oscillatory~field)},
\label{hallpotential1}
\end{equation}
where $\rho_{||}$ is an appropriate average of
$\rho_{xx}$ and $\rho_{yy}$ and standard symbols
for the components of the resistivity tensor are used.

Under the conditions of almost complete penetration by the oscillatory
field, the phase of $V_{\rm H}$ in Eqn.~\ref{hallpotential1}
with respect to $\tilde{H}$
is the same as that of a voltage induced in an open loop (i.e. $\pi/2$).
However, if significant screening occurs, $V_{\rm H}$
will also contain a component in quadrature to this. 
Using standard complex notation, we write
\begin{equation}
V_{\rm H} = V^\prime_{\rm H} - {\rm i}V^{\prime\prime}_{\rm H},
\end{equation}
where $V^\prime_{\rm H}$ and $V^{\prime\prime}_{\rm H}$ 
are the dissipative and reactive (quadrature) components
of $V_{\rm H}$ respectively.
In our experiment, phase-sensitive detection 
techniques allowed $V^\prime_{\rm H}$
and $V^{\prime\prime}_{\rm H}$ to be simultaneously
recorded for analysis purposes; the signal
from a multiturn pick-up coil mounted 
close to the sample was used to define
the {\it real} phase.  
The approximate distribution of the current was
further determined by placing pairs 
of contacts arranged between the
outer and upper surfaces of the sample of \khg ~of 
volume~$\approx 1$~mm$^3$
shown in Fig.~\ref{samplephoto}.  
While the finite size of
100~$\pm$~50~$\mu$m of the electrical 
contacts applied using graphite
paint limited the spatial resolution of 
the experiment, the contacts do enable
one to distinguish between different models.  
During these
experiments, temperatures down to 0.5~K were 
achieved using a plastic $^3$He refrigerator.

The orbital magnetic currents, 
${\bf j}_{\rm mag}$, were determined by
measuring the magnetic torque of a different 
$M=$~K sample to that shown in
Fig.~\ref{samplephoto} as well as a $M=$~Rb sample.
In each case, $\theta$, the angle between the
magnetic field ${\bf H}$ and the normal to the sample's
conducting planes was restricted to small angles
($\theta \leq 20^\circ$); this
avoided complications due to magnetic torque 
interaction~\cite{singleton1}.
The torque was measured capacitively by
attaching the sample to a $5~\mu$m-thick 
phosphor-bronze
cantilever that forms one of the plates of a capacitor.
Electrical contacts applied to the sample
confirmed that ${\bf j}_{\rm mag}$
continued to persist after the electrical currents dissipate.
In these experiments, temperatures down 
to 50~mK were provided by a dilution refrigerator.
\section{Hall potential measurements}
\label{honoldstuff}
\begin{figure}[htbp]
\centering
\includegraphics[width=8cm]{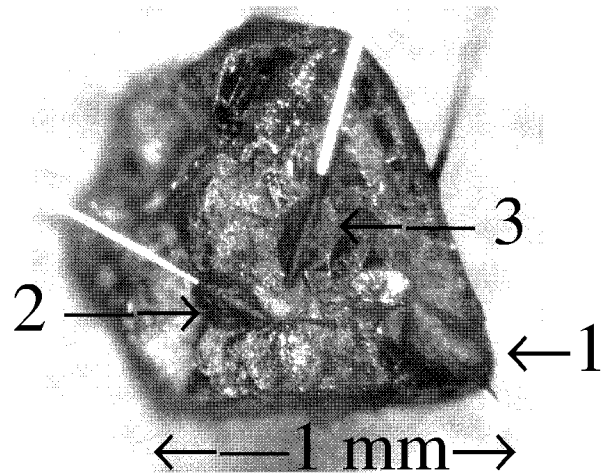}
\caption{A photograph of the sample 
of \khg used in the measurements
of the Hall potential, showing three 
contacts on the upper surface of
the sample.  Contacts on the side are not in focus.}
\label{samplephoto}
\end{figure}

\begin{figure}[htbp]
\centering
\includegraphics[width=8cm]{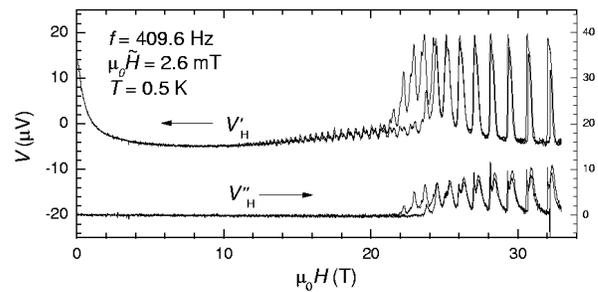}
\caption{Examples of the real and imaginary Hall potentials
$V^\prime_{\rm H}$ and $V^{\prime\prime}_{\rm H}$ measured
with the third pair of contacts on the sample shown in
Fig.~\ref{samplephoto}. An oscillatory
applied field with $\mu_0\tilde{H}\approx 2.6$~mT rms 
and $\omega/2\pi =409.6$~Hz was used; $T\sim$~0.5~K.
Arrows indicate the axis corresponding to each data set.}
\label{Hallpotential}
\end{figure}

Fig.~\ref{Hallpotential} shows typical
measured Hall potentials $V^\prime_{\rm H}$ and 
$V^{\prime\prime}_{\rm H}$
due to an oscillatory field; in this case 
$\mu_0 \tilde{H} = 2.6$~mT,  $\omega/2\pi=409.6$~Hz 
and the temperature of the
\khg ~sample ($A \approx 1.1$~mm$^2$) was 0.5~K.
The $V_{\rm H}$ data in Fig.~\ref{Hallpotential}
correspond to the contact 
labelled 3 in Fig.~\ref{samplephoto};
as the contact on the upper
face of the sample was very near to its geometrical 
center, we expect that
the Hall potential will be close to the maximum value possible.
Note that there was a small field-independent inductive pick-up
contribution $V_{\rm pu}$
to $V_{\rm H}'$ due to the 
open-loop area ($\ltsim 1$~mm$^2$)
between the contacts. By comparing measured $V^\prime_{\rm H}$
values at several different fields, it is possible to
infer that $V_{\rm pu} \approx -4.7~\mu$V.

In the discussion that follows, we first attempt to analyse the
data of Fig~\ref{Hallpotential} in terms of conventional
bulk currents within the
{\it whole} of the \khg ~sample; the analysis uses 
conductivity and resistivity tensor elements which
do not vary spatially.
We shall find that whilst this conventional analysis
describes the behavior of $V^{\prime}_{\rm H}$,
albeit with some remarkably low values of $\rho_{||}$,
it is unable to account  
for the variation of $V^{\prime\prime}_{\rm H}$
with $H$.
\subsection{Analysis in terms of bulk currents}
In a reasonably isotropic
quasi-two-dimensional metal, the 
components of the resistivity $(\rho)$ and conductivity $(\sigma)$
tensors are related by the expressions
\begin{equation}
\rho_{xy} = \frac{\sigma_{xy}}{\sigma_{||}^2+\sigma_{xy}^2},~~~~~
\rho_{||} = \frac{\sigma_{||}}{\sigma_{||}^2+\sigma_{xy}^2}.
\label{tensor}
\end{equation}
Thus, the ratio $\rho_{xy}/\rho_{||}$ in Eqn.~\ref{hallpotential1}
is equivalent to $\sigma_{xy}/\sigma_{||}$;
it is therefore likely that the asymptotic 
variation of $V^\prime_{\rm H}$ at low fields is
due to the divergence of 
\begin{equation}
\sigma_{xy} = \varrho_{\rm 2D}/\mu_0 H,
\label{oldhall}
\end{equation}
where $\varrho_{\rm 2D}$ is the charge density of the 2D holes,
as $H\rightarrow 0$.

As $H$ increases, $\rho_{xy}/\rho_{||}$ becomes smaller, leaving
$V_{\rm pu}$ as the dominant contribution to the measured
$V^\prime_{\rm H}$.
However, after the ``kink'' transition at $\mu_0 H \approx 23$~T, 
$V^\prime_{\rm H}$ undergoes a resurgence and the quadrature component
$V_{\rm H}''$ becomes significant for the first time. The maxima
in $V^{\prime}_{\rm H}$ occur at integral 
Landau level filling factors $\nu=F/\mu_0H$, where
$F$ is the magnetic quantum oscillation frequency;
at such fields, the
chemical potential $\mu$ is situated in a Landau gap.
The peak values $V^{\prime}_{\rm H}\approx 25~\mu$V 
(at $\mu_0 H \approx 25$~T) correspond to
$\rho_{xy}/\rho_{||} \approx  42$; 
the fact that $\rho_{xy} \gg \rho_{||}$
allows us to use 
$\rho_{xy} \approx 1/\sigma_{xy}=\mu_0H/\varrho_{\rm 2D}$ 
(see Eqn.~\ref{oldhall})
with $\varrho_{\rm 2D}/e=1.6 \times 10^{26}$~m$^{-3}$~\cite{singleton1} 
to extract
a minimum sample resistivity of 
$\rho_{||} \approx 2.4 \times 10^{-8}~\Omega$m
at the maxima close to $\mu_0H=30$~T.
Such resistivity values are characteristic of a
good metal at room temperature and are not too dissimilar from
the results of AC susceptibility experiments on 
\khg~\cite{harrison12}.

To examine the apparent bulk resistivity components in more
detail, we rearrange Eqs.~\ref{tensor}
in terms of $\rho_{||}/\rho_{xy}$ and $\sigma_{xy}$
to yield
\begin{equation}\label{rhoxy}
    \rho_{xy}=\frac{1}{\sigma_{xy}\big[1+\big(\frac{\rho_\|}{\rho_{xy}}
    \big)^2\big]}
\end{equation}
and
\begin{equation}\label{rhoxx}
    \rho_\|=\frac{\big(\frac{\rho_\|}{\rho_{xy}}
    \big)}{\sigma_{xy}\big[1+\big(\frac{\rho_\|}{\rho_{xy}}
    \big)^2\big]}.
\end{equation}
These equations can then be used along with
Eqns.~\ref{hallpotential1} and~\ref{oldhall}
to convert the experimental
values of $V^{\prime}_{\rm H}$ to
apparent bulk in-plane resistivity components. 

\begin{figure}[htbp]
\centering
\includegraphics[width=8cm]{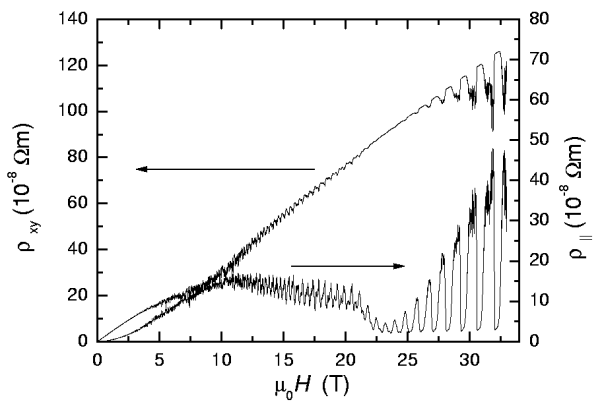}
\caption{The bulk resistivity tensor elements $\rho_{xy}$ and
$\rho_\|$ estimated from $V^\prime_{\rm H}$ using 
Eqs.~\ref{rhoxy} and \ref{rhoxx} ($T = 0.50$~K). 
Arrows indicate the axis corresponding to each data set.}
\label{resistivity}
\end{figure}

The results of this procedure 
are shown in  Fig.~\ref{resistivity}; the
linear increase of $\rho_{xy}$ with $H$ and quadratic increase of
$\rho_\|$ with $H$ in at low magnetic fields
show that the method gives a
behavior in accord with expectations.
At higher fields, the deduced values of $\rho_{xy}$
behave in a similar manner to earlier direct 
measurements of oscillations in the Hall
resistivity~\cite{harrison7}.  However, it is
interesting to note that the in-plane resistivity is lower in
Fig.~\ref{resistivity} than that 
obtained from four-terminal methods~\cite{harrison7}. 
Owing to non-uniform current flow associated with sample
imperfections,  direct measurements of the in-plane resistivity
are often contaminated by the inter-plane component $\rho_{zz}$,
which is several orders larger in magnitude~\cite{harrison7}.

\begin{figure}[htbp]
\centering
\includegraphics[width=8cm]{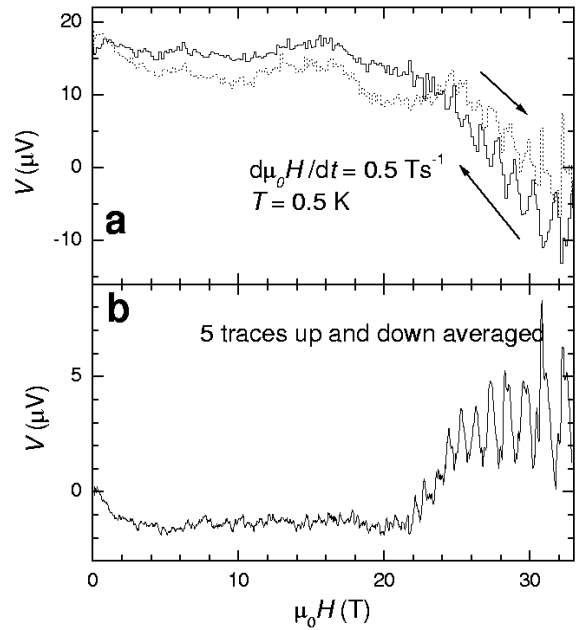}
\caption{An example of the Hall potential 
$V_{\rm H}$ measured for
the third pair of contacts in 
a slowly varying magnetic field
(sweep rate $\partial\mu_0H/\partial t=$ 0.5~T$^{-1}$; $T=0.50$~K).
(a) shows raw data for a single sweep, 
with arrows indicating the sweep direction.
No filtering has been applied. 
The background non oscillatory
component could be caused by variations
in the contact potentials with
magnetic field. (b) The difference 
$(V_{\rm H,up}-V_{\rm H,down})/2$
between rising- and falling-magnetic-field 
data averaged over 5 sweeps.}
\label{dcHallpotential}
\end{figure}

The unusually large value of $\rho_{xy}/\rho_\|=$~42 
and consequent
low value of $\rho_\|$ estimated from the oscillatory field 
measurements are confirmed by 
setting $\tilde{H}=0$ and by sweeping the quasistatic field.
Data obtained using $\mu_0 \partial H/\partial t = 0.5~$Ts$^{-1}$
are shown in Figure~\ref{dcHallpotential} for both rising and falling
magnetic fields. Insertion of this rate of change of field into
Eqn.~\ref{hallpotential2},  along with the value
$\rho_{xy}/\rho_\| \approx 42$ obtained in the oscillatory-field
experiments (see above), yields $V_{\rm H}\approx 1.8~\mu$V at
$\mu_0H \approx 25$~T, in good agreement with the data shown in
Fig.~\ref{dcHallpotential}.
However, the magnitude of the Hall potential continues to increase
with increasing field, rising to a value of $V_{\rm H}\approx 8~\mu$V
at around 32~T (Fig.~\ref{dcHallpotential}).  This yields
$\rho_{xy}/\rho_\|=180$, corresponding to
$\rho_\|\approx 6 \times 10^{-9}~\Omega$m, which is several
times lower than the resistivity of room-temperature copper.  Such
values for these parameters would not be unusual for a two-dimensional
electron gas exhibiting the quantum Hall effect~\cite{chakraborty1}.
However, they are unusual for a bulk organic 
metal far from the quantum limit;
for example $\nu\approx 21$ at 32~T.

\subsection{Analysis in terms of edge currents}
Thus far, a conventional treatment employing
uniform conductivity tensor elements 
$\sigma_\|$ and $\sigma_{xy}$
has given an {\it apparently} satisfactory explanation
of the Hall potential measured in swept fields
and the real component of the Hall potential $V^\prime_{\rm H}$
observed in oscillatory fields, 
albeit with extremely low values of the bulk resistivity.
However, this simple model fails on considering 
the imaginary component $V^{\prime\prime}_{\rm H}$ 
which appears within the high magnetic field regime in
Fig.~\ref{Hallpotential}. 
Were \khg ~a conventional metal, then in order to 
have a quadrature component
of order 30~\% of the in-phase component at 
$\omega/2\pi = 409.6$~Hz, the
skin-depth must be comparable to or shorter 
than the effective radius $r_0=\sqrt{A/\pi}\sim$~590~$\mu$m
of the sample~\cite{abrikosov1}.  If we
estimate the skin-depth $\delta=\sqrt{2\rho_\|/ \mu_0\omega}$ 
from the above resistivity value
($\rho_\|\sim$~2.4~$\times$~10$^{-8}$~$\Omega$m), 
however, we obtain a
value $\delta=$~3.9~mm that is 
significantly greater than the sample size.  
The presence of a large finite 
$V^{\prime\prime}_{\rm H}$ term, therefore, cannot be
easily explained by a simple conductivity tensor model.

\begin{figure}[htbp]
\centering
\includegraphics[width=8cm]{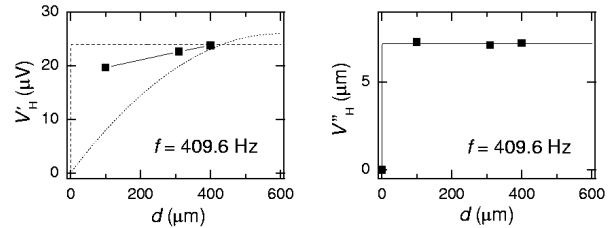}
\caption{The Hall potential difference 
between the edge and a voltage
probe situated a distance $d$ 
inside the upper surface where $d=$~100, 310 and 400~$\mu$m 
for contacts 1, 2 and 3 respectively, with
$T=$~0.5~K, $f=$~409.6~Hz and 
$\mu_0\tilde{H}=$~2.6~mT. (a) shows the
normal Hall voltage $V^\prime_{\rm H}$.  
The dotted line shows the
voltage distribution expected according to 
Eq.~\ref{Halldistribution1} while the 
dashed line shows the voltage
distribution expected for an exponential variation
of the Hall potential, $V_{\rm H} \propto \exp(-d/\lambda)$.  
(b) shows the reactive Hall 
voltage $V^{\prime\prime}_{\rm H}$, with the 
solid line depicting $V_{\rm H} \propto \exp(-d/\lambda)$ 
with $\lambda\ll$~100~$\mu$m.}
\label{Halldistribution}
\end{figure}

A further indication of the fact that the induced currents
do not obey a conventional bulk mechanism is given by
using the contact arrangments shown in
Fig.~\ref{samplephoto} to explore the dependence of
$V^\prime_{\rm H}$ and $V^{\prime\prime}_{\rm H}$
on the distance $d$ of the contact on the upper surface of the
sample from the edge.  
For a spatially homogeneous metallic system in
a slowly time-varying magnetic field, it is trivial to show that
the current density and
concomitant Hall potential should vary quadratically;
\begin{equation}
\label{Halldistribution1}
    V_{\rm H}(d)=
    \frac{d(2r_0-d)}{r_0^2}~V_{\rm H, max},
\end{equation}
where $V_{\rm H, max}$ is the total potential difference between the
outer surface and geometric centre of the sample. 
This form occurs because  the size of the area loop
element susceptible to induced voltages 
increases parabolically with
the distance from the geometric centre of the sample.  
On comparing the detected Hall potentials 
for the three different contact
arrangements shown in Fig.~\ref{samplephoto}, 
we find that $V^\prime_{\rm H}$ at 
$\mu_0 H \approx 25$~T in Fig.~\ref{Halldistribution}
varies more slowly with $d$ than 
predicted by Eq.~\ref{Halldistribution1}.  
In the case of the reactive component, no
discernable dependence of 
$V^{\prime\prime}_{\rm H}$ on $d$ is seen. 
This type of behavior shows that the Hall potential
difference (and therefore also the current) is heavily weighted
towards the sample edge, with the effect being particularly
pronounced for the imaginary component.

\begin{figure}[htbp]
\centering
\includegraphics[width=8cm]{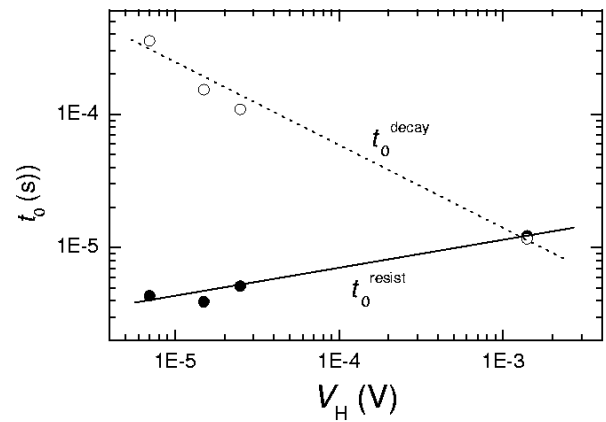}
\caption{A comparison of the decay times 
$t_0^{\rm decay}$ and $t_0^{\rm resist}$.
All estimates were made for $\mu_0H\approx 25$~T 
and $T=0.5$~K, with the data points corresponding 
to excitation frequencies $\omega/2\pi=51.25$,
208.6, 409.6 and 8809~Hz and oscillating
field amplitudes $\mu_0\tilde{H}=6.51$, 3.90, 2.62 
and 3.75~mT.}
\label{times}
\end{figure}

Currents carried at or close to the edge are 
expected to occur in quantum-Hall-effect systems for
which $\sigma_\|\rightarrow$~0 at integral 
Landau-level filling factors.
These currents are either carried within edge
channels~\cite{chakraborty1}, giving rise 
to a kind of chiral Fermi
liquid~\cite{balents1}, 
or are bulk states near the edge in which
electrostatic forces locally deplete 
(or enhance) the charge carrier
density~\cite{macdonald1}.  
Were a situation where $\sigma_\|\rightarrow$~0 
realised in \khg, perfect screening would
give rise to a Hall potential that is 
entirely reactive, even in the
limit as $\omega \rightarrow 0$.  Experimentally, 
$|V^{\prime\prime}_{\rm H}/V^\prime_{\rm H}| \ltsim~1$,
implying two things:
first, $\sigma_\|\neq0$ within the bulk, 
and second, the
partial screening that takes place has 
to occur within a region around
the perimeter of the sample that 
is thinner than the bulk skin-depth.

In the Appendix, we show that a system comprising a CDW
coexisting with quasi-two-dimensional Landau levels can lead
to very effective screening currents within a distance
$\lambda = \sqrt{m/2\mu_0 e \varrho_{\rm 2D}}$ of the sample surface.
Although the finite size of the electrical 
contacts in Fig.~\ref{samplephoto}
($100\pm 20~\mu$m in diameter), 
limits the resolution to
which we can analyse the Hall voltage distribution in
Fig.~\ref{Halldistribution}, we can, nevertheless,
perform some simple consistency checks on this model.
We propose that all of the current flows within a distance
$\lambda$ of the surface of the sample, and that this
current flow encounters a characteristic (very low)
resistivity $\rho_\lambda$.
The system is like an $LR$ (inductance-resistance) circuit,
with 
\begin{equation}
L=\frac{\alpha \mu_0\pi r_0^2}{l}
\label{inductance}
\end{equation}
and
\begin{equation}
R=\frac{2\pi r_0\rho_\lambda}{\lambda l},
\label{resistance}
\end{equation}
where $l$ is the sample height, $r_0=\sqrt{A/\pi}$ is
its effective radius and $\alpha$ is a so-called
Nagaoka parameter, a numerical factor $\sim 1$
that depends on $l/r_0$~\cite{page}. 
For our sample, $\alpha \approx 0.5$.
In a magnetic field swept at a slow uniform rate, this treatment yields
\begin{equation}
V_{\rm H} = \frac{r_0\lambda}{2}\left(\frac{\rho_{xy}}{\rho_\lambda}\right)
\mu_0 \frac{\partial H}{\partial t} + V_{\rm H0}{\rm e}^{-t/t_0}
\label{sweptedge}
\end{equation}
(c.f. Eqn.~\ref{hallpotential2})
whereas for a sinusoidally-oscillating field we obtain
\begin{equation}
V_{\rm H} = V^\prime_{\rm H}-{\rm i}V^{\prime\prime}_{\rm H}
=\frac{r_0\lambda}{2}\left(\frac{\rho_{xy}}{\rho_\lambda}\right)
\frac{\mu_0 \omega \tilde{H}}{1+\omega^2 t^2_0}(1-{\rm i}\omega t_0).
\label{oscedge}
\end{equation}
Here 
\begin{equation}
t_0 = L/R =\alpha \mu_0 r_0 \lambda/2\rho_\lambda
\label{tdefn}
\end{equation}
is a characteristic inductive decay time and $V_{\rm H0}{\rm e}^{-t/t_0}$
represents transient solutions due to rapid changes in 
$\partial H/\partial t$ and/or $R$.

We first compare Eqn.~\ref{sweptedge} with the Hall
potential of 8~$\mu$V observed in Fig.~\ref{dcHallpotential}
at $\mu_0H\approx 32$~T, obtaining 
$\rho_\lambda/\lambda\approx 23~\mu\Omega$,
or $\rho_\lambda\approx 9 \times 10^{-12}~\Omega$m, using
$\lambda\approx 400$~nm~(see Appendix).  Inserting this into the
semiclassical Drude expression 
$\rho_\lambda=m/e\varrho_{\rm 2D}\tau$, we
obtain a scattering time of $\tau\approx 50$~ns, 
or, a mean free path 
$\Lambda=v_{\rm F}\tau\approx 5$~mm 
(where we have obtained the Fermi velocity from
$v_{\rm F}=\sqrt{2\hbar eF}/m$).  
The fact that $\Lambda$ is
comparable to the sample perimeter 
of 4~mm could be consistent with
ballistic transport.  
The ballistic transport regime, for which finite
values of $\sigma_\|$ can no longer be locally 
defined, is one of the
essential preconditions of the edge-weighted current
model (see Appendix).

Two separate estimates of $t_0$ can be made.
First, we can rearrange Eqns.~\ref{oscedge} and \ref{tdefn} 
to yield $t_0 = 
\alpha V^\prime_{\rm H}(1+\omega^2t^2_0)/\rho_{xy}\omega\tilde{H}$.
In the limit of low frequency, 
$\omega^2t_0^2 \ll 1$, so that this can be written
\begin{equation}
t^{\rm resist}_0 = \frac{\alpha V^\prime_{\rm H}}{\rho_{xy}\omega\tilde{H}},
\end{equation}
where the superscript $^{\rm resist}$ denotes that 
this estimate of the characteristic time is derived from the dissipative
(i.e. resistive) component of $V_{\rm H}$.
An alternative estimate can be made by comparing the
resistive and reactive parts of $V_{\rm H}$:
\begin{equation}
t^{\rm decay}_0=\frac{|V^\prime_{\rm H}|}{\omega|V^{\prime\prime}_{\rm H}|},
\end{equation}
where the superscript $^{\rm decay}$ denotes that this is analogous to the
technique used to find the decay time of conventional $LR$ circuits.

Fig.~\ref{times} shows $t^{\rm resist}_0$ and
$t^{\rm decay}_0$ deduced from experimental data
($\mu_0 H \approx 25$~T, temperature = 0.5~K).
If the sample surface layer functioned as a conventional
(but high-conductivity) metal, we should expect the two times
to be the same. However, the divergence 
between $t^{\rm decay}_0$ and $t^{\rm resist}_0$ 
at low frequencies, or low values of $V_{\rm H}$, is
unconventional.  This suggests that the sample behaves
resistively when attempting to 
drive a current through it, but once
established, the currents have a 
tendency to last longer than
expected for a conventional metal.  
Their lifetime 
$t^{\rm decay}_0$ becomes particularly long 
at low values of the Hall
potential $V_{\rm H}$ observed in 
slowly-varying magnetic fields.

In rapidly-varying magnetic fields, $t^{\rm decay}_0$
converges with $t^{\rm resist}_0$ at $V_{\rm H}\approx 1$~mV.
This is similar to the saturation value of $V_{\rm H}$ observed
in ms-duration pulsed magnetic fields~\cite{honold2,honold3}. 
 Attempts to explain this saturation
in terms of the breakdown of the quantum Hall 
effect have been largely
unsuccessful owing to the 
enormous magnitude of the electric field
that is required.  
An electric field that is concentrated towards the
sample edge within a distance $\lambda$~(see Appendix), 
however, increases the probability of Zener tunneling.
Moreover, the increased size of the unit
cell in the CDW phase compared to 
the normal metallic phase makes
Zener tunneling across the CDW gap 
more likely than Landau level
tunneling.  In such a model for the 
magnitude of the threshold electric field 
$E_{\rm Z}=\varepsilon_{\rm g}^2/ea\varepsilon_{\rm F}$
(where $\varepsilon_{\rm g}=2\Psi$ is the CDW gap and
$\varepsilon_{\rm F}$ is the Fermi energy), 
the lattice periodicity
$a=2\pi/Q\approx\pi/k_{\rm F}$ occurs in the 
denominator.  Assuming an exponential variation of the 
electric field where $V_{\rm H}=\lambda E_{\rm Z}$, we obtain 
$\lambda E_{\rm Z}=4\Psi^2\lambda/\pi e\hbar v_{\rm F}\approx 3$~meV 
for $2\Psi=1$~meV, which is comparable to the
saturation observed in pulsed magnetic field
studies~\cite{honold2,honold3}.

Should normal scattering processes be inhibited according to the
screening model described in the Appendix, Zener
tunneling would provide a natural means for the decay of the
currents, given the potentially large values of the 
electric field concentrated close to the sample edge.
\section{magnetic torque measurements}
\label{torquetorque}
\begin{figure}[htbp]
\centering
\includegraphics[width=8cm]{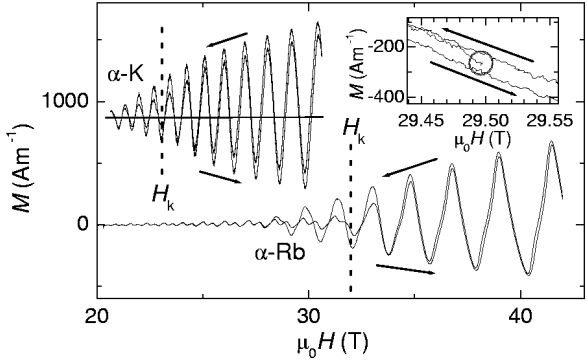}
\caption{The magnetization $M$ of 
samples of the $M=$~K ($\theta \approx 19.5^{\circ}$)
and Rb ($\theta \approx 5^{\circ}$) salts
measured on both rising and falling 
fields at $T=$~50~mK for $M=$~K
and $T=$~0.45~K for $M=$~Rb made using the 
magnetic torque method.
Arrows indicate the direction of sweep.  
For the $M=$~Rb salt, at
fields below approximately 29~T, 
the CDW$_0$ phase is stable.  
The region between 29~T and 33~T appears 
to be disturbed since $H_{\rm k}$
is different between rising and falling fields.  
At fields above
approximately 33~T, the transition into 
the CDW$_x$ phase is complete.
The de~Haas-van~Alphen oscillations 
are the same for rising and
falling fields, apart from an 
relative offset 2$\Delta M$.  For the
$M=$~K salt, the magnetic torque shows 
similar features to that
measured in Reference~\cite{harrison2}. 
The inset shows the
magnetization measured after a 
field cool (as indicated in Fig.~1).
The magnetization starts out at the 
value indicated by the large
center-dot circle, after which 
the field is swept down to
$\mu_0H<$~29~T followed by an upsweep 
to $\mu_0H>$~30~T.}
\label{torque}
\end{figure}

The Hall potential associated with free 
currents ${\bf j}_{\rm free}$
becomes difficult to detect 
for magnetic field sweep rates much less
than 0.5~Ts$^{-1}$.  In contrast, magnetic currents 
${\bf j}_{\rm mag}$ continue to be present 
for an arbitrarily long period of time
after the magnetic field sweep is stopped~\cite{harrison2}.  
Fig.~\ref{torque} shows examples
of the magnetization of 
$\alpha$-(BEDT-TTF)$_2M$Hg(SCN)$_4$ (for
$M=$~K and Rb) estimated from the magnetic torque. 
Magnetic fields of $\mu_0 H \approx 23$~T and 
$\mu_0 H\approx 32$~T are
required to access the CDW$_x$ phase for $M=$~K and Rb 
respectively~\cite{singleton1};
in this phase, hysteresis similar to that originally
obtained by Christ {\it et al.}~\cite{harrison2,christ2} is observed.

\begin{figure}[htbp]
\centering
\includegraphics[width=8cm]{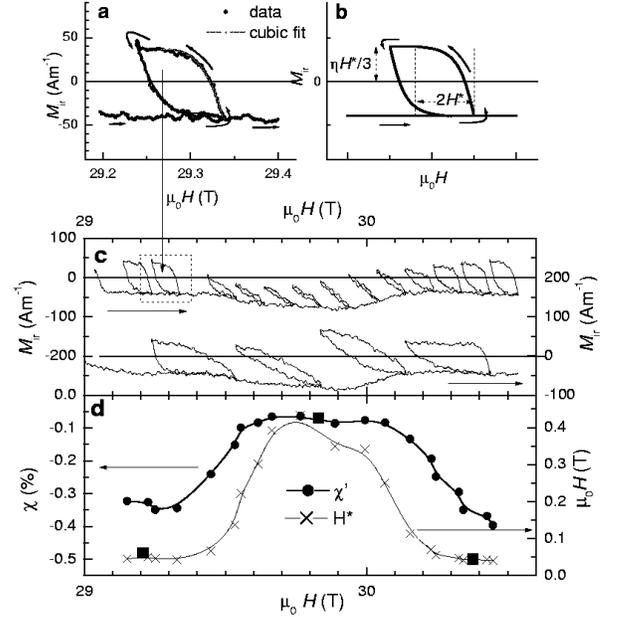}
\caption{Evidence for persistent currents 
in \khg.  ~(a) An
example of a loop in the non-equilibrium 
component of the magnetization $\Delta M$ measured 
as the field is swept up to
29.33~T then momentarily down to 29.24~T 
before being resumed.  Arrows
indicate the change in the locus of 
$\Delta M$ versus $H$.  The
solid lines show the results of 
fits of $\Delta M$ to Eq.~\ref{beanman}.
(b) A model hysteresis loop calculated
according to Eq.~\ref{beanman}
(i.e. the Bean model for a long cylinder with its 
axis parallel to $H$~\cite{bean1,poole1}), 
showing the theoretical saturation value of
$\Delta M_z$ and the reversal state.  
(c) A series of $\Delta M$
versus $H$ loops like that in (a) measured over 
an extended interval
of field (top), and (bottom), 
a similar measurement but where the
field interval over which the sweep 
direction is reversed is increased.  
Arrows indicate the appropriate axes.  (d) Estimates of
$\Delta\chi$ from fits of Eq.~\ref{beanman} 
to the many loops shown in
(c) (full circles) together with those 
(large full squares) calculated
according to Eq.~\ref{irrev} (see also Ref.~\cite{harrison11}).
Estimates of the coercion field $H^\ast$ (x-symbols) are also shown.
Arrows indicate the appropriate axes.}
\label{loops}
\end{figure}

The magnetization ${\bf M}$ of quasi-two-dimensional metals
such as \khg  ~is largely directed normal
to the conducting layers.  A torque
$\tau=\frac{1}{2}\mu_0M_zH\sin2\theta$ per unit volume 
therefore occurs when
there is a finite angle $\theta$ between ${\bf M}$ and 
and the applied field ${\bf H}$~\cite{singleton1}.
Sawtooth-like de~Haas-van~Alphen oscillations 
(originating from the
closed section of Fermi surface that survives the CDW 
formation~\cite{singleton1}) account for
the larger, reversible contribution to ${\bf M}$.  These
oscillations occur in all metals with 
closed sections of Fermi surface
at low temperatures and result from changes in the equilibrium
populations of orbitally quantized Landau levels as $H$ is
swept~\cite{shoenberg1,maniv1,harrison13}.  
The waveform of the
oscillations in the $M=$~K salt 
at high magnetic fields has received
much theoretical attention and is well
understood~\cite{maniv1,harrison13}.

By contrast, the contribution which we attribute to ${\bf j}_{\rm mag}$
is irreversible, and depends on how the sample is driven to a
particular field and temperature.
An example of this is shown in the inset to Fig.~\ref{torque};
if the sample is cooled to 0.5~K at fixed field, 
the magnetization attains
the value surrounded by the circle. On sweeping the
field down for the first time, the magnetization rises to meet
the path which will be followed by subsequent downsweeps.
Moreover, it proves to be impossible to go back to the encircled
value of ${\bf M}$ by merely changing the field; 
data recorded on subsequent
upsweeps fall below it, whilst downsweep data lie above it
(Fig.~\ref{torque}, inset). One can only recapture the
field-cooled value by warming the sample well above the CDW$_x$
phase and cooling again in the presence of a steady field.
  
Thus, sweeping the field produces 
an irreversible contribution to the magnetization
that manifests itself as an offset $2\Delta M$ between 
rising and falling field data in Fig~\ref{torque}.
Fig.~\ref{loops} shows $\Delta M$ for a \khg ~sample; 
values have been extracted by subtracting
the reversible de~Haas-van~Alphen effect contribution
(obtained by averaging full up and down sweeps) from the
raw magnetization data to leave just the irreversible component.
The magnetic field in Fig.~\ref{loops}a is first swept slowly up to
29.34~T, then down to 29.24~T, after which the up sweep is resumed.
Each time the magnetic field sweep 
direction is reversed, a finite
interval in field $\Delta H>2H^\ast$ 
(where $H^\ast$ is the coercion
field) is required in order for $\Delta M$ to reach a saturation
value that opposes the direction of sweep.

The Appendix describes how the CDW system produces
a diamagnetic contribution to the magnetization 
that acts to partially screen the
applied magnetic field (see Eq.~\ref{irrev} and the discussion
following it). However, this contribution saturates when the force
on the CDW, ${\bf F}$, resulting from a spatially-varying free energy,
exceeds the pinning force, ${\bf F}_{\rm p}$. Elastic energy considerations
(see Appendix) show that the region in which $|{\bf F}| > |{\bf F}_{\rm p}|$
propagates inwards from the sample surface as the magnetic field changes,
so that the volume-fraction of the sample able to
contribute to the diamagnetic screening decreases.
This is analogous to the scenario in the Bean model
of type-II superconductors~\cite{harrison11,bean1,poole1},
yielding, under the approximation of a cylindrical sample,
an irreversible susceptibility 
\begin{equation}
\Delta\chi\equiv \frac{\partial \Delta M}{\partial H}
=\Delta\chi_0f(\Delta H)=\Delta\chi_0[1-\frac{\Delta H}{2H^\ast}]^2.  
\label{beanbean}
\end{equation}
and an irreversible magnetization
\begin{equation}
\Delta M=(\Delta\chi_0\frac{H^\ast}{3})(2[1-\frac{\Delta H}{2H^\ast}]^3-1).
\label{beanman}
\end{equation}
Fig.~\ref{loops}b shows the predicted behavior of the irreversible
component of $M$ according to Eq.~\ref{beanman}, and 
Fig.~\ref{loops}a shows a fit of the model to the experimental data.
The fact that Eq.~\ref{beanman} reproduces the
experimental result indicates that the volume
fraction of the sample $f(\Delta H)$ that is able to respond with an
irreversible susceptibility $\Delta\chi_0$ declines in a simple
quadratic manner with $\Delta H$ (see Eq.~\ref{beanbean}), 
behaviour typical
of critical-state models such as that of 
Bean~\cite{harrison11,bean1,poole1};
here, $f(\Delta H)$ is the cross-sectional
area of the remaining non-critical portion of the sample.  

As the magnetization varies throughout the sample, Maxwell's
relation ${\bf j}_{\rm mag}=\nabla\times{\bf M}$~\cite{bleaney} 
implies that a magnetic-field-induced
critical state always involves currents.
However, although a critical-state model provides an excellent
fit to the experimental data in Fig.~\ref{loops}a, it is important to
note that there are two essential properties of the currents giving
rise to the loops in Fig.~\ref{loops} that distinguish them from
those in superconductors.  
First, the magnitude of the non-equilibrium
susceptibility $\Delta\chi_0\approx 3\times10^{-3}$
(obtained from the fits) departs significantly 
from the ideal diamagnetic value
$\Delta\chi_0\approx-1$ observed in superconductors.  
This implies
that the magnetic currents in 
$\alpha$-(BEDT-TTF)$_2M$Hg(SCN)$_4$
cannot be considered as screening currents 
in the same sense as those
in superconductors.  Because the magnetic currents 
screen only a part
of the magnetic field, the effective coercion field 
$H^\ast\approx j_{\rm c}r_0/\Delta\chi_0$ is rather large.  
Second, the value of
$\Delta\chi_0$ varies strongly as a function of $H$ in
Figs.~\ref{loops}c and \ref{loops}d, a property that would be very
difficult to explain in terms of superconductivity, but which
is an intrinsic feature of the model of a CDW coexisting with
well-defined Landau levels discussed in the Appendix. 

Fig.~\ref{loops}c shows a series of loops 
in $\Delta M_z$ versus $H$
like that in Fig.~\ref{loops}a, 
from which $\Delta\chi_0$ is extracted
at various values of $H$ in Fig.~\ref{loops}d.  
The magnitude of
$\Delta\chi_0$ is low and of order 
5$\times$10$^{-4}$ when $\mu$ is
situated in the middle of a Landau level 
(at half-integral filling
factors), but increases almost ten fold 
when $\mu$ is between Landau
levels (at integral filling factors).  
The theoretical limit for
$\Delta\chi_0$ becomes equivalent 
to that of a superconductor (i.e.
$\Delta\chi_0=-1$) only in that case of 
an ideal sample at integral
filling factors in which the 
quantum lifetime is infinite (see Eq.~\ref{irrev}
in the Appendix, remembering that 
demagnetising factors~\cite{bleaney}
must be taken into account once the susceptibility
becomes very large).

When pinning of the CDW occurs, 
the non-equilibrium susceptibility
$\Delta\chi_0$ provides a measure 
of how quickly the CDW departs from
equilibrium as the magnetic field is swept.  
This occurs more rapidly
at integral Landau level filling factors 
where the density of states
is lowest, enabling $\mu$ to jump quickly between 
Landau levels~\cite{harrison11}.  
Pinning of the
CDW prevents the equilibrium redistribution 
of carriers between the
bands, causing this jump to become 
much more abrupt.  This can account
the origin of the strong variations in $\Delta\chi_0$ in
Fig.~\ref{loops}d, as shown by the model calculations (see Appendix).

\begin{figure}[htbp]
\centering
\includegraphics[width=8cm]{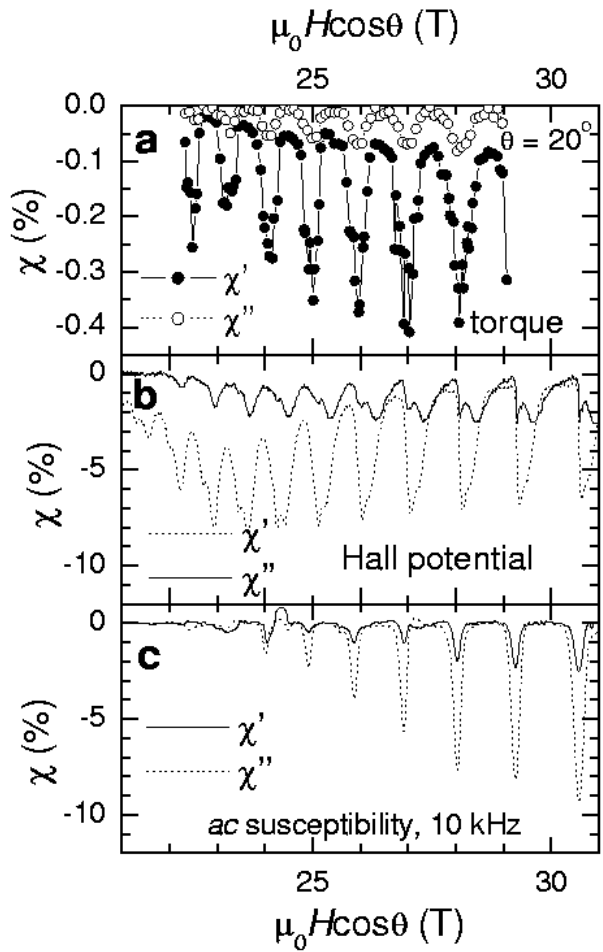}
\caption{A comparison of ac susceptibilities 
(or equivalent ac
susceptibilities) made on 
$\alpha$-(BEDT-TTF)$_2$KHg(SCN)$_4$ using
different methods.  In (a) this is determined by 
fitting Eq.~\ref{beanman}
to many hysteresis loops over an extended interval 
in magnetic
field.  The imaginery component is 
determined from the area of the
loop divided by its width.  
In (b), the effective susceptibility is
obtained by equating 
$\chi^\prime=2\pi V_{\rm H}^{\prime\prime}/\rho_{xy}\mu_0\tilde{H}$, 
where $V_{\rm H}^{\prime\prime}$ is the 
reactive component of the Hall potential. 
The imaginary susceptibility is given by 
$\chi^\prime=2\pi V_{\rm H}^{\prime}/\rho_{xy}\mu_0\tilde{H}$.  
Here, the susceptibility is
estimated from the data in Fig.~3. 
(c) is the published ac
susceptibility measured on another 
sample~\cite{harrison12}.}
\label{comparison}
\end{figure}

\section{discussion and conclusion}
\label{discussion}
Before concluding this paper, it is 
intructive to compare effective ac
susceptibilities of $\alpha$-(BEDT-TTF)$_2$KHg(SCN)$_4$ determined
using different methods in Fig.~\ref{comparison}.  In
Fig.~\ref{comparison}a, 
$\chi^\prime\equiv\Delta\chi_0$ is obtained by
fitting the Bean model (as described above- see Eq.~\ref{beanman}) 
to many hysteresis loops over an extended 
region of magnetic field.  The
imaginary component $\chi^{\prime\prime}$, 
which accounts for losses,
is obtained from the loop areas.  
In Fig.~\ref{comparison}b, the
susceptibility is estimated using
\begin{equation}\label{susceptibility}
    \chi^\prime-i\chi^{\prime\prime}=\frac{-2\pi i}{\rho_{xy}\tilde{H}}
    (V_{\rm H}^\prime-iV_{\rm H}^{\prime\prime}).
\end{equation}

The most clear aspect of Fig.~\ref{comparison} 
is that the magnitudes of
the real and imaginary susceptibilitiies 
due to free currents 
${\bf j}_{\rm free}$, estimated from the ac Hall potential, 
agree more
closely with the high frequency 
ac susceptibility measurements made in
Ref.~\cite{harrison12} (shown in Fig.~\ref{comparison}c) 
than with the magnetic torque measurements.  
This indicates that the ac
susceptibility made at frequencies 
$\omega/2\pi \gtrsim50$~Hz consists mostly of
free currents ${\bf j}_{\rm free}$, which 
are shown (above, and in the Appendix) 
to be confined to the sample edges.  
The data in Fig.~\ref{times}
shows that ${\bf j}_{\rm free}$ can persist 
for times as long as 
$t_0^{\rm decay}\gtrsim 0.5$~ms for sufficiently 
small values of the
Hall potential $V_{\rm H}\lesssim 7~\mu$V in a sample of
$\sim 1$~mm$^2$ cross-section at $\mu_0H\approx 25$~T and 
$T\approx0.5$~K.
In spite of the fact that the corresponding 
susceptibility in
Fig.~\ref{comparison}b is much larger than 
that for the torque
measurements in Fig.~\ref{comparison}a, 
the magnetization $M_{\rm free}=V_{\rm H}/\rho_{xy}\approx 7$~Am$^{-1}$ 
due to free currents is actually smaller.  
Free currents therefore have insufficient duration
to contribute a significant amount to the 
irreversible steady magnetic
torque in Fig.~\ref{loops}.  
By contrast, magnetic currents observed in the
torque experiments last for an indefinite period of
time~\cite{harrison2}.

The combined induction of free and magnetic 
currents ${\bf j}_{\rm free}+{\bf j}_{\rm mag}$ 
can therefore be understood as follows: for
small changes in magnetic field 
$\mu_0\Delta H\ll 10$~mT, the greatest
contribution to the susceptibility 
initially originates from free
currents at the surface that 
attempt to screen changes in magnetic
flux density from within the bulk.  
These currents quickly saturate,
however, as dissipation sets in, 
possibly assisted by Zener tunneling.
At that point, changes in magnetic flux 
density enter the bulk which
cause the Landau level structure 
and chemical potential to change,
causing the CDW to depart from equilibrium, 
as described in the Appendix.  
The stored energy increases quadratically with
$\Delta H$, eventually causing the 
CDW to collapse at the edges,
initiating the critical state.

In conclusion, Hall potential and magnetic torque
measurements on $\alpha$-(BEDT-TTF)$_2M$Hg(SCN)$_4$ 
($M=$~K, Rb) show that two types of 
screening currents occur within
the high-field, low-temperature CDW$_x$ phase
in response to changing magnetic fields.
The first, which gives rise to the induced Hall 
potential, is a free current (${\bf j}_{\rm free}$),
weighted mostly towards the edge of the sample. 
The time constant for the decay of these currents
is longer than that expected from the sample
resistivity. The second component of the current 
appears to be magnetic (${\bf j}_{\rm mag}$), 
in that it is a microscopic, quasi-orbital
bulk effect; it is evenly distributed within
the sample upon saturation.
A simple model (Appendix), describing a new
type of quantum fluid comprising a CDW coexisting
with a two-dimensional Fermi-surface pocket,
is able to account for the origins 
of the currents.
Taken together, these findings are able
to reconcile the body of experimental 
evidence~\cite{harrison2,harrison6,hill1,harrison7,honold1,
honold2,harrison8,harrison9,mielke1} 
which had previously been interpreted
in terms of the quantum Hall effect~\cite{singleton1} 
or superconductivity~\cite{harrison8}.

\section*{Acknowledgements}
This work is supported by US
Department of Energy (DoE) under
grant LDRD-DR 20030084 and by the 
UK Engineering and Physical Sciences Research Council (EPSRC).
The National High Magnetic Field Laboratory
is supported by DoE,  the National
Science Foundation (NSF) and the State of Florida.
We thank Stan Tozer and Alessandro Narduzzo for experimental
assistance, and acknowledge stimulating discussions with
James Brooks, Jason Lashley and Albert Migliori.
\section*{Appendix: Screening mechanisms in \khg}
We treat a simplified version of the  \khg 
~bandstructure~\cite{singleton1}, comprising a 1D 
electron band and a two-dimensional
(2D) hole 
band with dispersions 
$\varepsilon_{\rm 1D}=\hbar v_{\rm F}|k_x-k_{\rm F}|$ and 
$\varepsilon_{\rm 2D}=\varepsilon_{\rm F}-\hbar^2(|k_x-k_X|^2+|k_y-k_Y|^2)/2m$ 
respectively;
here $m$ is the effective mass, 
$v_{\rm F}$ is the Fermi velocity of
the 1D band, and $k_X$ and $k_Y$ define the 
centre of the 2D hole pocket.  
Each of these bands intersects the Fermi energy
($\varepsilon_{\rm F}$), giving rise to the 
simplified Fermi surface shown in the inset to Fig.~1.

The simplest scenario to consider is that
where only the 1D Fermi-surface section is 
subject to CDW formation,
giving rise to a gap $2\Psi$ in its density of 
electronic states,
while the 2D hole section remains ungapped.  
Under equilibrium conditions, the average 
volume charge densities $\bar{\varrho}_{\rm 1D}$ and
$\bar{\varrho}_{\rm 2D}$ associated with the 
1D and 2D Fermi-surface sections respectively 
would be subject to the conservation equation
$\bar{\varrho}_{\rm 1D}+\bar{\varrho}_{\rm 2D}+\varrho_{\rm bg}=0$, 
where $\varrho_{\rm bg}$ is the
density of charge due to the ionic cores.  
However, below we consider 
slight local deviations $\Delta \varrho_{\rm 1D}$ 
and $\Delta \varrho_{\rm 2D}$ in the
charge density from the equilibrium values 
(i.e. the total local
charge densities associated with the 1D and 2D 
Fermi-surface sections become 
$\varrho_{\rm 1D}=\Delta \varrho_{\rm 1D}+\bar{\varrho}_{\rm 1D}$ and 
$\varrho_{\rm 2D}= \Delta \varrho_{\rm 2D}+\bar{\varrho}_{\rm 2D}$ respectively); 
because of the presence of the CDW, the overall 
conservation equation need no longer hold {\it locally}. 
However, the charge densities must
obey Poisson's equation
\begin{equation}
-\epsilon \nabla^2 V({\bf r}) = \Delta\varrho_{\rm 2D} + \Delta \varrho_{\rm 1D},
\label{monsieurfish}
\end{equation}
where $V$ is the electrostatic potential and 
$\epsilon$ is the permittivity.
We are at liberty to set the origin of potential; we choose $V=0$
in the absence of spatial charge variations.
Equation~\ref{monsieurfish} implies that the presence of local 
charge-density variations will lead to an 
in-plane, spatially-varying electric field
${\bf E}=-\nabla V$. 

We now introduce the magnetic flux density ${\bf B}_0$ applied along $z$
(i.e. ${\bf B} = (0,0,B) \equiv B{\bf \hat{z}}$, 
$\perp$ to the 2D $(x,y)$ planes).
This has two effects; first, the crossed ${\bf E}$ and ${\bf B}$ fields force
the 2D hole wavefunction centres to
drift at a velocity ${\bf E}\times{\bf B}_0/B_0^2$.
This leads to a spatially-varying 
in-plane current density, which, provided that the scattering
rate is rather small (e.g. under conditions of ballistic or quasi-ballistic
transport), can be written as
 ${\bf j} \approx \varrho_{\rm 2D}{\bf E} \times{\bf B}_0/B_0^2$;
note that the relationship becomes exact in the total absence of scattering.
Integration of Maxwell's fourth equation 
($\nabla \times {\bf H} = {\bf j}$~\cite{bleaney}) shows that
this current produces an additional contribution 
\begin{equation}
\Delta{\bf B} = -\mu_0 \varrho_{\rm 2D}(V/B_0){\bf \hat{z}}
\label{fromage}
\end{equation}
to the magnetic flux density 
parallel to $z$, which then becomes ${\bf B}= {\bf B}_0+\Delta {\bf B}$.
A second effect of ${\bf B}$  is to produce Landau quantisation of the 2D
holes; this is usually dealt with using the Landau 
gauge ${\bf A}=(0,Bx,0)$~\cite{book}.
After some manipulation~\cite{book}, the Schr\"{o}dinger equation
for the in-plane wavefunctions $\psi$ 
and eigenenergies $E$ may be written as
$\left(-\frac{\hbar^2}{2m}\frac{\partial^2}{\partial x^2} + \frac{1}{2} m \omega^2_{\rm c}(x-x_0)^2 \right) \psi
=E\psi $.
Here, the effect of the crystalline potential 
has been taken into account by the effective mass $m$;
the cyclotron frequency is $\omega_{\rm c} = qB/m$, with $q$ the charge;
$x_0$ represents the wavefunction guiding-centre
coordinate. 

The in-plane electric field also
modifies the Landau-level energies~\cite{macdonald1}.  
To quantify this,
we consider the case $V({\bf r})=V(x)$, {\it i.e.} the
potential varies only in the $x$ direction 
(variation in an arbitrary
direction in the $x,y$ plane is reintroduced below).
If $V(x)$ varies slowly over lengthscales $\sim x_0$,
it can be expanded about $x_0$ in a Taylor's series
$V(x)=V(x_0)+(x-x_0)V'(x-x_0)+\frac{1}{2}(x-x_0)^2V''(x_0)+.....$,
where the primes indicate differentiation with respect to $x$.  
After a little algebra, the Schr\"{o}dinger equation becomes
\begin{equation}
 -\frac{\hbar^2}{2m}\frac{{\rm d}^2\psi}{{\rm d}x^2} +
(\frac{1}{2}m(\omega^2_{\rm c} +\frac{q}{m}V''(x_0))(x-x_1)^2)\psi =E\psi, 
\label{maddog2} 
\end{equation} 
where $x_1$ is a constant 
(absorbing terms in $x_0$, $V(x_0)$ and $V'(x_0)$). 

The chief effect of $V(x)$ comes from $V"(x_0)$;
this gives a modified Landau-level spectrum, $(n+\frac{1}{2})\hbar \omega$,
with $n$ an integer and
\begin{equation}
\omega = \omega_{\rm c}\left(1+\frac{qV"}{2m \omega_{\rm c}^2}\right)^{\frac{1}{2}}
\approx \omega_{\rm c0}\left(1 + \frac{\Delta B}{B_0} 
+ \frac{V"}{2\omega_{\rm c0} B_0}\right),
\label{maddog3}
\end{equation}
where $\omega_{\rm c0}=qB_{\rm 0}/m$ and only terms
to leading order in $\Delta B$ and $V"$ 
are retained in the right-hand bracket.
Reintroducing a potential variation in an arbitrary intraplane direction
merely changes $V"$ in Eqn.~\ref{maddog3} to $\nabla^2V$.
 
In the absence of potential variations, 
the number of states per unit volume per Landau
level is $D=D_0= m\omega_{\rm c0}/\pi \hbar c$, 
where $c$ is the layer separation 
in the $z$ direction; the introduction of 
varying $V$ (and consequent $\Delta {\bf B}$) causes $\omega_{\rm c0}$
to change to $\omega$ (Eqn.~\ref{maddog3}), modifying this
to $D =m\omega/\pi \hbar c =D_0+\Delta D$, where
\begin{equation}
\Delta D = D_0\left(\frac{\Delta B}{B_0}+
\frac{\nabla^2 V}{2\omega_{\rm c0}B_0}\right).
\label{maddog4}
\end{equation}
The change in Landau-level degeneracy results in 
a change to the local 2D hole density,
\begin{equation}
\Delta\varrho_{\rm 2D}= \nu q \Delta D - 
\nu \beta q \Delta D- \beta q \bar{g}_{\rm 2D}\Delta \mu.
\label{gerrya}
\end{equation}
It is worth taking time to understand this equation 
fully, since it is the key to 
understanding the screening effects 
that are the point of this paper.
The first term of the right-hand side results from
the change in degeneracy of {\it all} of 
the occupied Landau levels;
positive $\nabla^2V({\bf r})$ or 
$\Delta {\bf B}$ gives a 
greater Landau-level 
degeneracy (Eqn.~\ref{maddog4}), 
producing an {\it increase} in the number of holes.  
The second term results from the energy shift of the $\nu$th
Landau level closest to the chemical potential 
$\mu$; in the high-field limit of well-resolved levels, 
the contribution from the tails of the Landau levels 
remote from $\mu$ can be ignored.  
As an increase in $\nabla^2V({\bf r})$ causes $\omega$ to increase
(Eqn.~\ref{maddog3}), this results in a slight 
{\it depopulation} of the Landau levels closest to $\mu$, 
acting to reduce $\varrho_{\rm 2D}$.
The dimensionless factor $\beta$ therefore depends 
critically on the position of $\mu$ amongst 
the Landau levels; i.e. it oscillates as a
function of $B$.  If $\mu$ is in the middle 
of the highest occupied Landau level 
(at half-integral Landau-level filling, $\nu=F/B$),
the density of states is large, so that 
the shift in energy of the Landau level has a 
relatively large effect on $\varrho_{\rm 2D}$;
in this situation, 
$\beta_{\rm half}\approx 2\omega_{\rm c}\tau/\pi\gg 1$.  
On the other hand, if $\mu$ is directly between 
two Landau levels (at integral Landau-level filling 
$\nu=F/B$) then the density of states
is small and $\beta_{\rm int}\approx 4/\pi\omega_{\rm c}\tau \ll 1$.
This tends to zero in the case of 
a sample of pristine purity for
which the quantum lifetime $\tau\rightarrow\infty$~\cite{macdonald1}. 
The final term accounts for 
possible spatial variations $\Delta \mu$
in the chemical potential which are 
sustained by the CDW (see below);
here, $\bar{g}_{\rm 2D}\equiv \bar{\varrho}_{\rm 2D}m/\hbar F q^2$ 
is the mean value of the 2D density of states.

A shift in the chemical potential also affects the 1D carrier density;
\begin{equation}
\Delta \varrho_{\rm 1D} = -eg_{\rm 1D} \Delta \mu ,
\label{davidt}
\end{equation}
where $g_{\rm 1D}$ is the 1D density of states.
We can now reformulate Eqn.~\ref{gerrya} by
substituting from Eqns.~\ref{monsieurfish}, 
\ref{fromage}, \ref{maddog4} and \ref{davidt}
to yield
\begin{equation}
\frac{1}{2 \omega_{\rm c0}}\nabla^2 V+ \Delta B
=\frac{m}{\hbar e \nu} \left(\frac{\beta + \eta}{1-\beta}\right)\Delta \mu .
\label{everything}
\end{equation}
We have retained terms up to first order in 
small quantities, omitted $\epsilon \nabla^2 V$ from
Eqn.~\ref{monsieurfish} because it is $\sim 10^4$ times smaller
than the other term in $\nabla^2V$ 
and substituted $q=+e$ for the hole 
charge;\footnote{Note that the result depends
only on $q/m$, and so is independent of whether 
we consider the 2D Fermi surface as
a hole pocket with positive $m$ and $q$ or an electron 
pocket with negative $m$ and $q$.} 
here, $\eta = g_{\rm 1D}/g_{\rm 2D}$.

\begin{figure}[htbp]
\centering
\includegraphics[width=8cm]{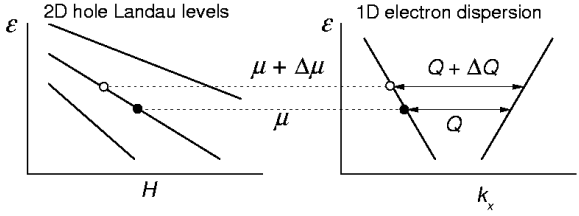}
\caption{A model depicting the affect of a change
$\Delta\mu\propto\Delta N_{\rm 1D}$ in the chemical potential $\mu$ on
the 2D hole Landau levels and the 1D electron dispersions (in this
case at half-integer filling), giving rise to a direct relationship
between $\Delta N_{\rm 1D}$, $\Delta\mu$ and $\Delta Q$ and
consequently between $\Delta Q$ and $H$.}
\label{mechanism}
\end{figure}

As long as the 1D band remains metallic, $\Delta \mu$ is
unconstrained.  Scattering facilitates the 
transfer of current between positive and
negative variations in the charge density,
causing them to dissipate rapidly.  
The stable result is when $\nabla^2 V= 0$, 
causing Eqn.~\ref{everything} to 
become a simple proportionality between $\Delta \mu$
and $\Delta B$; i.e. a variation in the 
chemical potential due to a change in 
magnetic field~\cite{harrison13}. 

By forming a CDW on the 1D bands, 
the system can benefit in two ways.
First, by the opening of a CDW gap $2\Psi$, the system has the
potential to become partly gapped at $\mu$.  
Second, by forming a quantum-coherent state,
the 1D band resists the spontaneous exchange of
charge between the two bands.  
As with a superconductor, the quantum state of
a CDW is locally defined by a phase $\phi$, which defines the
phase of the charge modulation that oscillates 
in one direction ($x$)
on a lengthscale $2\pi/|{\bf Q}|$ much shorter than the
cyclotron length. Any shift in the chemical 
potential in Eqn.~\ref{everything} 
will correspond to a gradient in phase
(see Fig.~\ref{mechanism});
\begin{equation}
\Delta \mu = 
\frac{\hbar v_{\rm F}}{2} \frac{\partial  \phi}{\partial x}    
\label{phase}
\end{equation}
The proportionality between 
$\partial \phi /\partial x$ and $\Delta \mu$
leads to two qualitatively 
different solutions to Eqn.~\ref{everything}.

\subsection*{Within the bulk of the sample}
In the bulk of the sample, the transport is
dissipative ($\nabla^2V \rightarrow 0$)
and the CDW is able to adjust its phase within certain limits
to accommodate currents caused by a non-uniform potential.
Eq.~\ref{everything} therefore becomes
\begin{equation}
\frac{\Delta \mu}{\Delta B} = 
\frac{\hbar e \nu}{m}\left(\frac{1-\beta}{\beta + \eta}\right).
\label{dong1}
\end{equation}
Thus, there is a direct proportionality between
local variations in the magnetic field and $\Delta \mu$.
As noted above (Eq.~\ref{phase}), a change $\Delta \mu$  results in
a shift in CDW phase, and, via~\cite{harrison11}
\begin{equation}
\frac{\partial \Delta Q}{\partial \mu} = 
\bar{Q}\frac{g_{\rm 1D}e}{|\varrho_{\rm 1D}|}
\end{equation}
to local adjustments of $Q$ away from its equilibrium value, $\bar{Q}$
(see Fig.~\ref{mechanism}).
These changes result in a concomitant shift in the free energy $\Phi$
of the system, which, following Ref.~\cite{harrison11}, can be written as
\begin{equation}
\Phi = -g_{\rm 1D}\left(\frac{\Psi^2}{2}-\Delta\mu^2\right)
+\delta g_{\rm 1D}\Delta \mu^2 + \Phi_{\rm 2D}(\varrho_{\rm 1D}).
\label{dong3}
\end{equation}
Here, $\Phi_{\rm 2D}$ is the free energy of the Landau-level system
(which depends at fixed field on the number of quasiparticles in
the 2D hole band, and hence on $\varrho_{\rm 1D}$)
and $\delta$ is a dimensionless parameter depending on the
position of the chemical potential amongst the Landau-levels;
it takes the limiting values~\cite{harrison11}
\begin{equation}
\delta = \frac{1}{2\eta \frac{\pi}{4}\omega_{\rm c} \tau}-
\frac{1+\eta}{2(\eta \frac{\pi}{4}\omega_{\rm c} \tau)^2}
\end{equation}
for integer Landau-level filling factors and $\delta = \frac{1}{2}$
for half-integer filling factors.

The effects described by Eqs.~\ref{dong1}-\ref{dong3} 
result in a relationship between $\Phi$ and $\Delta B$,
implying that there is a contribution $\Delta \chi$ from the CDW
to the overall susceptibility $\chi$, where
$\Delta \chi \equiv -\mu_0(\partial^2 \Phi /\partial (\Delta B)^2)$.
Using the identity
$(\partial \Phi /\partial (\Delta B))=
(\partial \Phi /\partial (\Delta \mu))\times 
(\partial (\Delta \mu)/\partial(\Delta B))$
in conjunction with Eqs.~\ref{dong1} 
and \ref{dong3} yields\footnote{Note that
this equation is the same as Eq.~4 of Ref.~\cite{harrison11}.}
\begin{equation}
\Delta \chi = -2g_{\rm 1D}\mu_0(1+\delta)
\left[\frac{\hbar e \nu}{m}\left(\frac{1-\beta}{\beta + \eta}\right)\right]^2.
\label{irrev}
\end{equation}
Thus, the non-equilibrium (i.e. spatially-varying) state of the
CDW thus provides a diamagnetic contribution to the overall susceptibility,
which resists changes in the magnetic field in the bulk of the sample;
note that the presence of the factors $\beta$ and $\delta$ indicate that
the effectiveness of this screening will oscillate as a function of
Landau-level filling factor $\nu$. However,
as we discuss in the next paragraph, Eq.~\ref{irrev} represents
the response of the sample only to infinitessimal changes in
magnetic field; the build up (and release) of elastic energy as the field
develops further leads to an irreversible change in the overall
sample susceptibility which eventually saturates.

The spatially-varying free energy
associated with the mechanism causing Eq.~\ref{irrev}
corresponds to a force ${\bf F}=-\nabla \Phi$; the non-equilibrium
excited state of the CDW will only persist if ${\bf F}$ remains
less than the CDW's pinning force ${\bf F}_{\rm p}$~\cite{harrison11}.
This produces two restrictions; first, because of
the differential relationship between ${\bf F}$
and $\Delta \mu$, $\Delta \mu$ must tend to zero at the sample surface,
or else ${\bf F}$ would be singular (and hence exceed ${\bf F}_{\rm p}$).
Second, once ${\bf F}$ develops with changing field to such an extent that the
CDW depins, the non-equilibrium build up of
$\Delta \mu$ must stop, 
preventing any further contribution from $\Delta \chi$.
These combine to ensure that the region in 
which $|{\bf F}| > |{\bf F}_{\rm p}|$
propagates inwards from the 
sample surface as the magnetic field changes,
so that the volume-fraction of the sample able to
contribute $\Delta \chi$ (Eq.~\ref{irrev})
to the overall volume-averaged
sample susceptibility $\chi$ becomes progressively smaller.
This is very similar to the 
Bean critical-state model~\cite{harrison11,bean1,poole1};
indeed, a common feature of {\it all} critical-state 
models is that motion
initiates at the surface 
because it is there that the potential energy
can be dissipated with the least amount of action.  
The CDW
begins to slide at the sample surface so that the restoring
force does not exceed the CDW depinning force.
Approximately linear gradients in $\Delta \varrho_{\rm 1D}$
and $\Delta Q$ occur because the work done 
by ${\bf F}$ is a linear function
of the departure $\Delta Q$ of the 
CDW from equilibrium~\cite{harrison11}.  
Currents result from changes in the
charge density $\varrho_{\rm 2D}$ associated with
the two-dimensional Fermi
surface section as the quasiparticles 
redistribute themselves so as to screen the
changes in the open Fermi surface section 
(see Eq.~\ref{monsieurfish}).
The variation in the orbital
magnetization of the closed Fermi-surface 
pocket $\Delta M$ with
$\Delta \varrho_{\rm 2D}$ then leads 
to linear gradients in $\Delta M$.
Maxwell's equations imply that a 
linear gradient in $\Delta M$,
sustained by pinning, corresponds 
to a critical current density 
${\bf j}_{\rm c}=\nabla\times(\Delta {\bf M})$, 
in the region where the CDW
is able to slide.

Therefore, although the mechanisms are very different, there are
close analogies between the situation here and flux-pinning
in superconductors.
It is therefore not surprising that
in Section~\ref{torquetorque}, we shall see that a simple Bean model
is able to fit the variation of sample susceptibility with
applied magnetic field.
\subsection*{Close to the sample surface}
We have already seen that $\Delta \mu \rightarrow 0$ as 
one approaches the sample surface because the
elastic energy cost of shifting the phase would otherwise
become too great.
Hence, Eq.~\ref{everything} assumes the form
\begin{equation}
\lambda^2 \nabla^2 V - V = 0~~{\rm with}~~\lambda= [m/(2 \mu_0 
e \bar{\varrho}_{\rm 2D})]^{\frac{1}{2}} ,
\label{gits}
\end{equation}
where we have substituted for $\Delta B$ using Eq.~\ref{fromage}.
Eq.~\ref{gits} defines a penetration depth 
$\lambda$ analogous to the London penetration depth in
superconductivity; it 
implies that deviations $V$ from the equilibrium value of the
electrostatic potential will be 
screened from the bulk of the sample.
Owing to the proportionality between 
$V$ and $\Delta B$ in Eqn.~\ref{fromage},
spatial variations in the magnetic field will also be screened.
Substituting values for \khg ~taken from Ref.~\cite{singleton1}
yields $\lambda \approx 400$~nm.

The existence of a finite $\nabla^2 V$
close to the surface implies quasi-ballistic transport.
By carrying the current
close to the edges, the total system can save energy
(in this case, elastic energy due to the non-equilibrium
arrangement of the CDW within the bulk of the sample), 
and any process
that saves energy also protects the quasipartices from ordinary
scattering events that would otherwise
give rise to dissipative bulk currents.  This
type of explanation for the variation in the current density,
electrostatic potential, electric field and perturbed charge carrier
density follows on from a model originally proposed by 
MacDonald {\it et al.}~\cite{macdonald1} for two-dimensional electron gas systems. 
Whereas for a single two-dimensional layer an exact solution can only
be obtained numerically, in bulk crystalline systems like
\khg, ~the 
above treatment shows that the variation in all of these
quantities becomes a simple exponential function.

\end{document}